%% file: osaka.tex
\documentclass{ws-p10x7}

\begin{document}
\input def.tex

\title{Precision measurements, extra generations and heavy neutrino }

\author{V.A.~Ilyin$^1$, M.~Maltoni$^2$, V.A.~Novikov$^3$, L.B.~Okun$^3$,
 A.N.~Rozanov$^{3,4}$  and M.I.~Vysotsky$^3$  }
\address{
$^1$  SINP Moscow State Univ., Moscow, Russia\\
$^2$  Instituto de F\'{\i}sica Corpuscular~--~C.S.I.C.,
Universitat de Val\`encia, Valencia, Spain\\ 
$^3$  ITEP, Moscow, Russia\\
$^4$  CPPM, IN2P3-CNRS, Univ. M\'editerran\'ee,
Marseilles, France\\
Presented by A.Rozanov, E-mail: rozanov@cppm.in2p3.fr 
}

\twocolumn[\maketitle\abstract{
The existence of extra chiral generations with all 
fermions heavier than $M_Z$
is strongly disfavored by the precision
electroweak data.
The exclusion of
one additional generation of heavy fermions in SUSY extension of Standard
Model is less forbidden
if chargino and
neutralino have low degenerate  masses with  $\Delta m \simeq 1$ GeV.
 However the data are fitted nicely even by a few extra
generations, if one allows neutral leptons to have masses close
to $50$ GeV. Such heavy neutrino can be searched in the reaction
$e^+ e^- \to N\bar{N}\gamma$ at LEP-200 with total final luminosity
of $2600 pb^{-1}$.
}]

\section{Introduction}\label{sec:intr}

 The straightforward generalization of the
Standard Model (SM) through inclusion of extra chiral generation(s) of
heavy fermions, quarks ($q=U,D$)  and leptons ($l=N,E$), is an example of
New Physics at high  energies
which does not decouple at ``low"
($\sim m_{Z}$) energies.
New particles contribute to physical observables through
self-energies  of vector and axial currents.
This gives 
corrections \cite{fourgen}
$\delta V_i$
to the functions $V_i (i=A,R,m)$ which determine  \cite{2}
the values of physical
observables (axial coupling $g_A$, the ratio $R=g_V/g_A$, and the ratio
$m_W/m_Z$).

 We consider the case of several lepton and quarks $SU(2)_L$ doublets and their
right-handed singlet
companions: $(UD)_L$, $U_R$, $D_R$, $(NE)_L$, $N_R$, $E_R$.
In what follows we will assume that the mixing among new generations and
the three existing ones is small, hence new fermions contribute only to
oblique corrections (vector boson self energies).

\section{LEPTOP fit to experimental data}\label{sec:LEPTOP}

We compare theoretical predictions for the case of
the presence of extra generations with
experimental data \cite{EWWG}
with the help of the code LEPTOP \cite{9XX}.
These experimental data are the latest updates presented at this conference
and they are well fitted by Standard Model.
We perform the four parameter ($m_t, m_H, \alpha_s, \bar{\alpha} $) fit
\footnote{The mass of Z-boson in the fit  was fixed to the latest experimental 
value $M_Z =  91.1875(21)$ GeV}
to
18 experimental observables.

\begin{table}[t]
\caption{LEPTOP fit of the precision observables.}\label{tab:observables}
\begin{tabular}{|l|l|l|r|}
\hline
Observ. & Exper.  & LEPTOP & Pull \\
           &  data   & fit    &      \\
\hline
$\Gamma_Z$ {\tiny[GeV]} &    2.4952(23) &  2.4964(16)  & -0.5   \\
$\sigma_h$ [nb] &   41.541(37)& 41.479(15)  & 1.7    \\
$R_l$ &   20.767(25)& 20.739(18)  & 1.1   \\
$A_{FB}^l$ & 0.0171(10)  &  0.0164(3)  & 0.7   \\
$A_{\tau}$ & 0.1439(42) &  0.1480(13)  & -1.0   \\
$A_e$ &  0.1498(48) &  0.1480(13) & 0.4  \\
$R_b$ &    0.2165(7) &   0.2157(1)  & 1.2   \\
$R_c$    &    0.1709(34)&   0.1723(1)  & -0.4   \\
$A_{FB}^b$  &    0.0990(20)&   0.1038(9)  & -2.4   \\
$A_{FB}^c$  &    0.0689(35)&   0.0742(7)  & -1.5   \\
$s_l^2$ {\tiny $(Q_{FB})$}   &    0.2321(10)  &   0.2314(2)  & 0.7   \\
\hline
$s_l^2$ {\tiny ($A_{LR}$)} & {\it 0.2310(3)}&  {\it  0.2314(2)}  & -1.5   \\
$A_b$ &  0.911(25) &  0.9349(1)  & -1.0   \\
$A_c$ &  0.630(26)&   0.6683(6)  & -1.5   \\
\hline
$m_W$ {\tiny [GeV]} & 80.434(37) & 80.397(23)  & 1.0   \\
$s_W^2$ {\tiny  ($\nu N$)} & 0.2255(21)& 0.2231(2) & 1.1 \\
\hline
$m_t$ {\tiny [GeV]}    & 174.3(5.1) &   { 174.0(4.2)} & 0.1\\
$m_H$ {\tiny [GeV]}    &   &  { $55^{+45}_{-26} $} &  \\
$\hat{\alpha}_s$ &           &  0.1183(27) &  \\
$\bar{\alpha}^{-1}$ & 128.88(9)           & 128.85(9) & 0.3  \\
{\small $\chi^2/n_{dof}$} & & 21.4/14 & \\
\hline
\end{tabular}
\end{table}
 
The fitted parameters
\footnote{ During this conference the new results on the electron-positron
  annihilation into hadrons in the range $\sqrt{s}=2-5$ GeV from BES \cite{BES}
  were released. With $\bar{\alpha}^{-1}=128.945(60)$ \cite{bolek}
  recalculated using these new BES results, we get from LEPTOP fit slighly
  higher prediction for the higgs mass
  $m_H =     78^{+53}_{-32}$ GeV,
  $m_t =     174.1(4.5)$ GeV,
  $\alpha_s =     0.1182(27)$,
  $\bar{\alpha}^{-1}=128.927(58)$ and 
  $\chi^2 / ndf =21.1/14 $.
}
together with the values of the predicted observables and their pulls
from the experimental data are given in the Table~\ref{tab:observables}.
Only the experimental value of the forward-backward assymetry in the 
Z decay into the pair of b-quarks $A_{FB}^b$ shows a hint for disagreement
with Standard Model.
 We take $m_D = 130$ GeV --
the lowest value allowed for the new quark mass from Tevatron search
\cite{13}  and take $m_U \ga m_D$.
As for the leptons from the extra generations, their masses are independent
parameters. To simplify the analyses we start with
 $m_N = m_U$, $m_E = m_D$.
Any value of higgs mass above $113.3$ GeV 
is allowed
\cite{higgs113} 
 in our fits, however
$\chi^2$ appears to be minimal for $m_H=113$ GeV.

\begin{figure}
\epsfxsize190pt
\figurebox{120pt}{190pt}{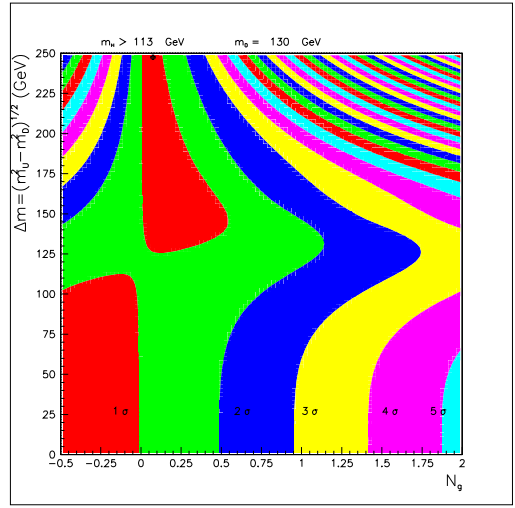}
\caption{
 Constraints on the number of extra generations $N_g$ and the mass
   difference in the extra generations $\Delta m$.
   The lowest allowed value  $m_D=130$ GeV from Tevatron search
   was used and $m_E=m_D$, $m_N=m_U$ was assumed.
}
\label{fig:mEmD}
\end{figure}

  In Figure \ref{fig:mEmD} the excluded
domains in coordinates ($N_g$, $\Delta m$) are shown 
(here $\Delta m = (m_U^2 - m_D^2)^{1/2}$).  Minimum
  of $\chi^2$ corresponds to $N_g =0.1$.  We see that one extra generation
corresponds to $2\sigma$ approximately.


We checked that similar bounds are valid for the general choice of heavy
masses of leptons and quarks. In particular we found that for 
 $m_N = m_D = 130$ GeV and $m_E = m_U$ one extra generation is excluded
at 1.5 $\sigma$ level, while for  $m_E = m_U = 130$ GeV and $m_N = m_D$
the limits are even stronger than in Fig.~\ref{fig:mEmD}.
So the extra generations are excluded by 
the electroweak precision data, if all
extra fermions are heavy: $m \ga m_Z$.


\section{Extra generations in case of SUSY}\label{sec:SUSY}

 When SUSY particles are heavy they decouple
 and the
same standard model exclusion plots shown in Fig.~\ref{fig:mEmD}  are valid.
One possible exception is a contribution of the third
generation squark doublet, enhanced by large stop-sbottom splitting. In this
way we get noticeable positive contributions to functions $V_i$  \cite{36,5},
which may help to compensate negative contributions of degenerate extra
generations.
\begin{figure}
\epsfxsize190pt
\figurebox{120pt}{190pt}{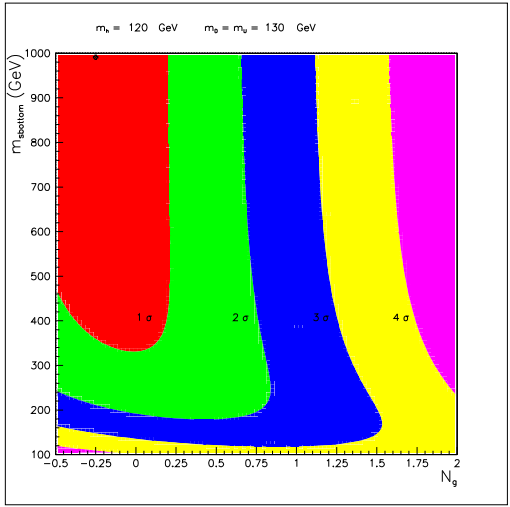}
\caption{
The 2-dimensional exclusion plot for the $N_g$
degenerate extra generations and the mass of sbottom $m_{\tilde{b}}$
in SUSY models and for the choice
  $m_D = m_U = m_E = m_N = 130$ GeV,
using $m_h =120 $ GeV, $m_{\tilde{g}} = 200$ GeV and assuming the absence
of $\tilde {t}_L - \tilde {t}_R$ mixing.
}
\label{fig:msb}
\end{figure}
 We analyze the simplest case of the absence of $\tilde {t}_L
-\tilde {t}_R $ mixing in Fig.~\ref{fig:msb}.  In this figure the case of degenerate extra
generations with common mass $130$ GeV is considered (contributions of
superpartners of new generations to $V_i$ are negligible since new up- and
down- particles are degenerate).  Exclusion plot is presented in coordinates
$(N_{g}, m_{sbottom})$.  We see that with inclusion of SUSY new
heavy generations are also disfavoured.

 Situation changes in case of light chargino and
neutralino. The latter are still not excluded - dedicated search at LEP II
still allows the existence of such  particles with masses as
low as 68 GeV 
(gaugino region with light sneutrino) \cite{charginoL3}
or 77 GeV 
(higgsino case) \cite{charginoDELPHI}
 if their mass difference is $\approx 1$ GeV.
Analytical formulas for corrections to the functions $V_i$ from
quasi degenerate chargino and neutralino were derived and analyzed in
\cite{14XX}. Corrections are big and this allows one to get lower bounds on
masses of chargino and neutralino:  $m_{\chi}>54$ GeV for the case of higgsino
domination and $m_{\chi}>61$ GeV for the case of wino domination at $95\%$
CL.

\begin{figure}
\epsfxsize190pt
\figurebox{120pt}{190pt}{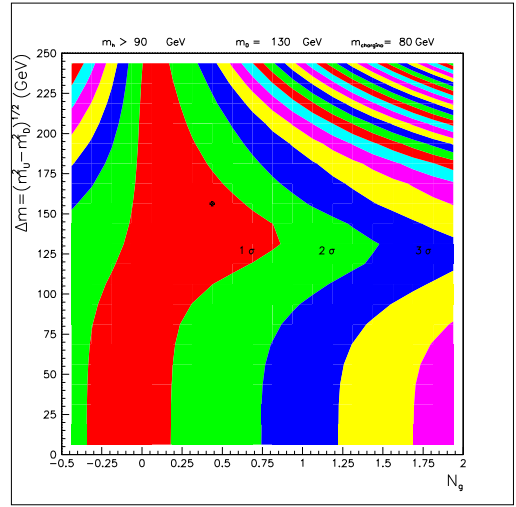}
\caption{
 Constraints on the number of extra generations $N_g$ and the mass
   difference in the extra generations $\Delta m$
in case of $80$ GeV
higgsino-dominated  quasi degenerate chargino and neutralino.
   The lowest allowed value  $m_D=130$ GeV from Tevatron search
   was used and $m_E=m_D$, $m_N=m_U$ was assumed.
}
\label{fig:degenerate}
\end{figure}

 Fig.~\ref{fig:degenerate} demonstrates how presence of chargino-neutralino pair (dominated
by higgsino) with mass $80$ GeV slightly relaxes the bounds shown on Fig.~\ref{fig:mEmD}. 
 We see that one
extra generation of heavy fermions is allowed within $1.5 \sigma$ domain
in case of the light chargino.

\section{Heavy neutrino with $m_N < m_Z$}\label{sec:heavy}

For particles with masses of the order of $m_Z/2$ oblique corrections
drastically differ from what we have for masses $\ga m_Z$. In particular,
renormalization of $Z$-boson wave function produces large negative
contribution to $V_A$.
 Quasi-stable neutral lepton $N$ should have the mass slightly above
 $m_Z/2$ to avoid increasing the invisible $Z$-width and it
should have the mixing angle with three known generations smaller
than $10^{-6}$ to avoid desintegration in the detector.
We consider new heavy neutrino  with Dirac mass and
we suppose that the Majorana mass of $N_R$ is negligible.
 From the analysis of the initial set of precision data
in papers \cite{2X,3X} it was found that
the existence of additional light fermions with masses $\approx 50$ GeV is
allowed.  Now analyzing all precision data
and using bounds from direct searches we conclude, that the only  presently
allowed light fermion is neutral lepton $N$.

\begin{figure}
\epsfxsize190pt
\figurebox{120pt}{190pt}{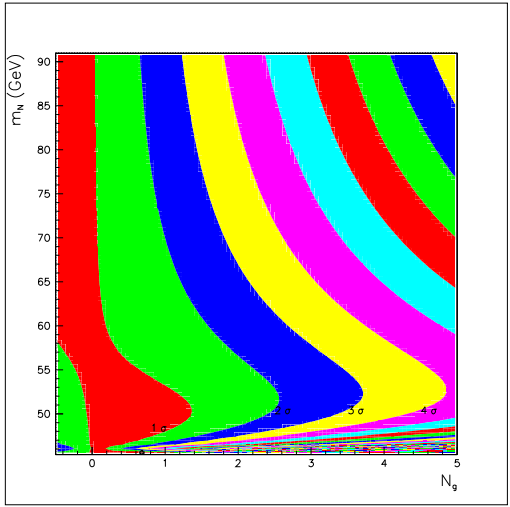}
\caption{
Constraints on the number of extra generations $N_g$ and the mass
   of the neutral heavy lepton $m_N$.
   The values  $m_U=220$ GeV, $m_D=200$ GeV, $M_E=100$ GeV
   were used.
}
\label{fig:mN}
\end{figure}

  As an example we take $m_U =
220$ GeV, $m_D = 200$ GeV, $m_E = 100$ GeV and draw exclusion plot in
coordinates ($m_N, N_g$), see Fig.~\ref{fig:mN}.
  From this  plot it is clear that for the case of fourth
generation with $m_N \approx 50$ GeV description of the data is not worse
than for the Standard Model and that even two new generations with $m_{N_1}
\approx m_{N_2} \approx 50$ GeV are allowed within $1.5 \sigma$.

\section{Possibility for the direct search of the 50 GeV heavy neutrino}\label{sec:search}

The direct search of the heavy neutrino is possible in
$e^+e^-$-annihilation into a pair of heavy neutrinos with the emission
of initial state bremsstrahlung photon
\begin{equation}
e^+ e^- \to \gamma + \; \mbox{\rm N} \bar{\mbox{\rm N}}
\label{eq:1}
\end{equation}
The main background is the production of the pairs of conventional
neutrinos with initial state bremsstrahlung photon
\begin{equation}
e^+ e^- \to \gamma + \; \nu_i \bar\nu_i
\label{eq:2}
\end{equation}
where $i = e, \mu, \tau$. These background neutrinos are produced
in decays of real and virtual $Z$. In case of  $\nu_e
\bar\nu_e$, two mechanisms contribute, through $s$-channel $Z$ boson
and from $t$-channel exchange of $W$ boson.
We calculated the signal and background distributions
and rates \cite{mn50} using CompHEP \cite{comphep} computer code.

\begin{figure}
\epsfxsize210pt
\figurebox{120pt}{210pt}{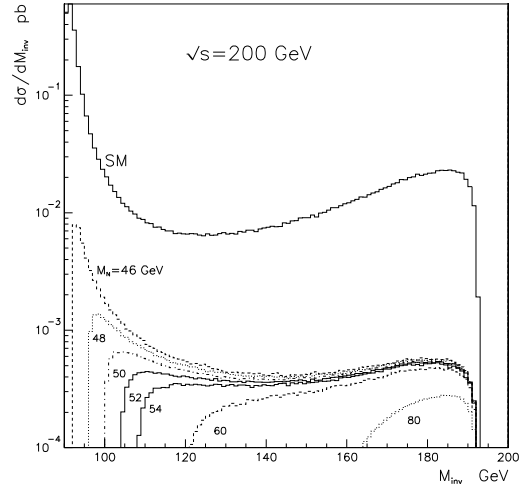}
\caption{
$d\sigma/d M_{inv}$ (in pb) for 
Standard Model and for the different values of $m_N$.
}
\label{fig:minv200}
\end{figure}

In Fig.~\ref{fig:minv200} the distribution on
``invisible" mass $M_{inv}$ (invariant mass of the neutrino 
pair)
is represented for SM background and the $N\bar N$ signal for $\sqrt{s}=200$
GeV and different values of $N$ masses, $M_N=46-100$ GeV. Here we applied
kinematical cuts on the photon polar angle and transverse momentum,
$|\cos\vartheta_\gamma|<0.95$ and $p_T^\gamma>0.0375\sqrt{s}$, being the
ALEPH selection criteria \cite{ALEPH}.
The photon detection efficiency 74\% is assumed.
For highest significance of the $N\bar N$ signal, evaluated as
$N_S/\sqrt{N_B}$, one should include whole interval on $M_{inv}$ allowed
kinematically, so we applied  $M_{inv}>2m_N$ cut.

\begin{figure}
\epsfxsize210pt
\figurebox{120pt}{210pt}{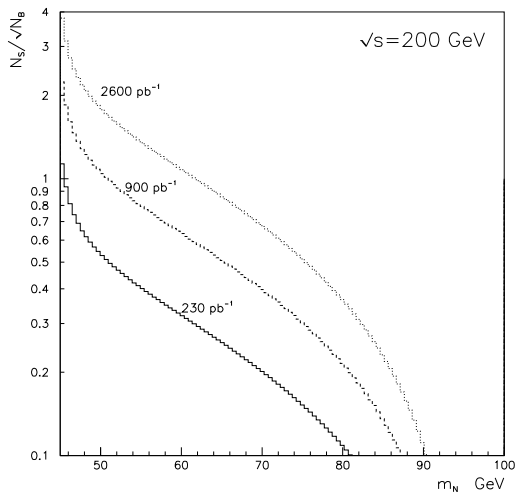}
\caption{
$N\bar N$ signal significances at LEP-2 at different statistics
as function of the neutrino mass. 
}
\label{fig:s-mN}
\end{figure}

On Fig.~\ref{fig:s-mN} the signal
significances
 are represented as a function
of $m_N$. One can derive that only the analysis based on combined data from
all four experiments both from 1997-1999 runs ($\sqrt{s}=182-202$ GeV) and
from the current run, in total $\sim 2600$ pb$^{-1}$, can exclude at 95\%
CL the interval of $N$ mass up to $\sim 50$ GeV.

Another possibility is to search for 50 GeV neutrino at the future
TESLA $e^+ - e^-$ electron-positron linear collider.
The increase in energy
leads to the decrease both of the signal and the background,
but it is 
compensated by the proposed increase of
luminosity
of 300 ${\rm fb}^{-1}$/year \cite{TESLA}.
Further advantage of the linear collider is the possibility to use
polarized beams.
 This is important in suppressing the cross section of $e^+ e^-
\to \nu_e\bar{\nu}_e\gamma$ as this reaction goes mainly through the
$t$-channel exchange of the $W$ boson.
 However, even without exploiting the beam
polarization the advantage of TESLA in the total number of events is
extremely important. Thus, Standard Model is expected to give approximately
0.3 million single photon events for $M_{inv}>100$ GeV while the number of
50 GeV neutrino pairs would be about 4000. 
%
Although the signal over background ratio is still small
(2.3-0.5\% for $m_N=45-100$ GeV correspondingly) the significance of the
signal is excellent, higher than 5 standard deviations for $m_N<60$ GeV.



\end{document}

%% file: def.tex
\def\la{\mathrel{\mathpalette\fun <}}
\def\ga{\mathrel{\mathpalette\fun >}}
\def\fun#1#2{\lower3.6pt\vbox{\baselineskip0pt\lineskip.9pt
\ialign{$\mathsurround=0pt#1\hfil##\hfil$\crcr#2\crcr\sim\crcr}}}
\newcommand{\nuanu}[1]{
 \stackrel{{\scriptscriptstyle (-)}}
                          {\nu}_{\hspace{-3pt}#1}}
\newcommand{\subn}[1]{\mbox{\scriptsize #1}}
\newcommand{\thetaW}{\theta_{\subn{W}}}
\newcommand{\sW}{s_{\subn{W}}}
\newcommand{\cW}{c_{\subn{W}}}
\newcommand{\cu}{c_{\subn{u}}}
\newcommand{\su}{s_{\subn{u}}}
\newcommand{\mZ}{m_{\subn{Z}}}
\newcommand{\mW}{m_{\subn{W}}}
\newcommand{\mWZ}{m_{\subn{W,Z}}}
\newcommand{\SigmaW}{\Sigma_{\subn{W}}}
\newcommand{\SigmaV}{\Sigma_{\subn{V}}}
\newcommand{\SigmapWZ}{\Sigma_{\subn{W,Z}}'}
\newcommand{\mH}{m_{\subn{H}}}
\newcommand{\mh}{m_{\subn{h}}}
\newcommand{\mA}{m_{\subn{A}}}
\newcommand{\mt}{m_{\subn{t}}}
\newcommand{\mSUSY}{m_{\subn{SUSY}}}
\newcommand{\ndf}{n_{\subn{d.o.f.}}}
\newcommand{\Ee}{E_{e}}
\newcommand{\pt}{p_{\subn{t}}}
\newcommand{\ppt}{p_{\subn{t}}^2}
\newcommand{\gV}{g_{\subn{V}}}
\newcommand{\gA}{g_{\subn{A}}}
\newcommand{\Vq}{\mbox{\rm V}_{\subn{q}}}
\newcommand{\Aq}{\mbox{\rm A}_{\subn{q}}}
\newcommand{\iq}{i_{\subn{q}}}
\newcommand{\gVq}{g_{\subn{$\Vq$}}}
\newcommand{\gAq}{g_{\subn{$\Aq$}}}
\newcommand{\RVq}{R_{\subn{$\Vq$}}}
\newcommand{\RAq}{R_{\subn{$\Aq$}}}
\newcommand{\Riq}{R_{\subn{$\iq$}}}
\newcommand{\gVe}{g_{\subn{V}}^{e}}
\newcommand{\gAe}{g_{\subn{A}}^{e}}
\newcommand{\gR}{g_{\subn{R}}}
\newcommand{\gL}{g_{\subn{L}}}
\newcommand{\gRL}{g_{\subn{R(L)}}}
\newcommand{\gLR}{g_{\subn{L(R)}}}
\newcommand{\thetaLR}{\theta_{\subn{LR}}}
\newcommand{\dSUSYLR}{\delta^{\subn{LR}}_{\subn{SUSY}}}
\newcommand{\dSUSY}{\delta_{\subn{SUSY}}}
\newcommand{\Eth}{\Ee\theta_{e}^2}
\newcommand{\nue}{\nu_{e}}
\newcommand{\anue}{\bar\nu_{e}}
\newcommand{\numu}{\nu_{\mu}}
\newcommand{\anumu}{\bar\nu_{\mu}}
\newcommand{\nutau}{\nu_{\tau}}
\newcommand{\anumue}{\bar\nu_{\mu}e}
\newcommand{\gn}{g^{\nu}}
\newcommand{\gnumu}{g^{\nu_{\mu}}}
\newcommand{\gnue}{g^{\nu_{e}}}
\newcommand{\gnutau}{g^{\nu_{tau}}}
\newcommand{\sbot}{\tilde{\mbox{\rm b}}}
\newcommand{\stot}{\tilde{\mbox{\rm t}}}
\newcommand{\stL}{\tilde{\mbox{\rm t}}_{\subn{L}}}
\newcommand{\stR}{\tilde{\mbox{\rm t}}_{\subn{R}}}
\newcommand{\stone}{\tilde{\mbox{\rm t}}_1}
\newcommand{\sttwo}{\tilde{\mbox{\rm t}}_2}
\newcommand{\sbL}{\sbot_{\subn{L}}}
\newcommand{\sbR}{\sbot_{\subn{R}}}
\newcommand{\msg}{m_{\subn{$\tilde{\mbox{\rm g}}$}}}
\newcommand{\msq}{m_{\subn{$\tilde{\mbox{\rm q}}$}}}
\newcommand{\mstL}{m_{\subn{\mbox{$\stL$}}}}
\newcommand{\mstR}{m_{\subn{\mbox{$\stR$}}}}
\newcommand{\mstone}{m_{\subn{\mbox{$\stone$}}}}
\newcommand{\msttwo}{m_{\subn{\mbox{$\sttwo$}}}}
\newcommand{\msb}{m_{\subn{\mbox{$\sbot$}}}}
\newcommand{\msbL}{m_{\subn{\mbox{$\sbL$}}}}
\newcommand{\msbR}{m_{\subn{\mbox{$\sbR$}}}}
\newcommand{\Vm}{V_{\subn{m}}}
\newcommand{\VR}{V_{\subn{R}}}
\newcommand{\VA}{V_{\subn{A}}}
\newcommand{\YL}{Y_{\subn{L}}}
\newcommand{\alphaW}{\alpha_{\subn{W}}}
\newcommand{\alphashat}{\hat{\alpha}_{\subn{s}}}
\newcommand{\alphabar}{\bar{\alpha}}
\newcommand{\unit}[1]{\, {\rm #1}}
\newcommand{\laeq}{\raisebox{-0.3ex}
                   {\footnotesize $\stackrel{\raisebox{-0.8ex}
                                              {$\textstyle <$}}{\sim}$}}
\newcommand{\diff}[2]{\frac{{\rm d}\,#1}{{\rm d}\,#2}}
\newcommand{\anu}{\bar\nu}
\newcommand{\andd}{\hspace*{1em} {\rm and}\ \hspace{1em}}
\newcommand{\with}{\hspace*{3em} {\rm with}\ \hspace{1em}}
\newcommand{\where}{\hspace*{3em} {\rm where} \hspace{1em}}